\documentclass[preprint, amsmath,amssymb,aps,prapplied,longbibliography]{revtex4-1}
\usepackage{graphicx}
\usepackage{dcolumn}
\usepackage{amsmath}
\usepackage{here}
\usepackage{bm}
\usepackage{color}
\usepackage[titletoc, title]{appendix}

\begin{document}

\title{
  Heavily Damped Precessional Switching 
  with Very Low Write-error Rate
  in Elliptical-cylinder Magnetic Tunnel Junction
}

\author{R. Matsumoto}\email{rie-matsumoto@aist.go.jp}
\author{S. Yuasa}
\author{H. Imamura}\email{h-imamura@aist.go.jp}
\affiliation{
  $^{1}$National Institute of Advanced Industrial Science and Technology (AIST),
  Research Center for Emerging Computing Technologies, Tsukuba, Ibaraki 305-8568, Japan
}

\date{\today}

\begin{abstract}
 Voltage-induced dynamic switching in magnetic tunnel junctions (MTJs) is a writing technique for voltage-controlled magnetoresistive random access memory (VCMRAM), which is expected to be an ultimate non-volatile memory with ultra-low power consumption. In conventional dynamic switching, the width of sub-nanosecond write voltage pulses must be precisely controlled to achieve a sufficiently low write-error rate (WER). 
 This very narrow tolerance of pulse width is the biggest technical difficulty in developing VCMRAM.
  Heavily damped precessional switching is a
  writing scheme for VCMRAM 
  with a substantially high tolerance of pulse width although the minimum WER has been much higher than that of conventional dynamic switching with an optimum pulse width. 
  In this study, we theoretically investigate the effect of MTJ shape and the direction of the applied magnetic field on the WER
  of heavily damped precessional switching.
  The results show that the WER in elliptical-cylinder MTJ can be several orders of magnitude smaller than that in usual circular-cylinder MTJ 
  when the external magnetic field is applied parallel to the minor axis of the ellipse.
  The reduction in WER is due to the fact that the demagnetization field narrows
  the component of the magnetization distribution perpendicular to the plane direction immediately before the voltage is applied.
\end{abstract}


\maketitle

\section{INTRODUCTION}
\label{introduction}
Voltage-controlled magnetoresistive random access memory (VCMRAM)
\cite{weisheit_electric_2007,
  maruyama_large_2009, duan_surface_2008,
  nakamura_giant_2009, tsujikawa_finite_2009,
  endo_electric-field_2010, shiota_induction_2012, shiota_pulse_2012,
  kanai_electric_2012,shiota_evaluation_2016, grezes_ultra-low_2016, shiota_reduction_2017,
  yamamoto_thermally_2018, yamamoto_improvement_2019}
has been attracting a great deal of attention as a low-power nonvolatile memory. 
The writing scheme of the VCMRAM is based on the voltage control of magnetic anisotropy (VCMA) at the interface between the MgO tunnel barrier and the free layer (FL) made of an Fe-based alloy such as Co-Fe in a magnetic tunnel junction (MTJ) \cite{yuasa_giant_2004, parkin_giant_2004, djayaprawira_230_2005} (see Fig. 1 (a)).  
When no voltage is applied to the MTJ, 
the magnetization in the FL is kept almost perpendicular to the plane direction by perpendicular magnetic anisotropy. The perpendicular magnetic  anisotropy can be reduced by applying a voltage pulse through the VCMA effect
\cite{weisheit_electric_2007, maruyama_large_2009, duan_surface_2008,
  nakamura_giant_2009, tsujikawa_finite_2009},
which induces magnetization precession around the external magnetic field
\cite{davies_anomalously_2019}. The magnetization switches if the voltage is turned off after half a precession period
\cite{endo_electric-field_2010, shiota_induction_2012, shiota_pulse_2012,
  kanai_electric_2012,shiota_evaluation_2016, grezes_ultra-low_2016, shiota_reduction_2017,yamamoto_thermally_2018, yamamoto_improvement_2019}. 
After the voltage pulse, the magnetization relaxes toward the equilibrium direction opposite to the initial direction and the switching completes. 
This is the conventional precessional-switching scheme of VCMRAM, which we refer to as dynamic precessional switching.

\begin{figure}[t]
  \includegraphics [width=1\columnwidth] {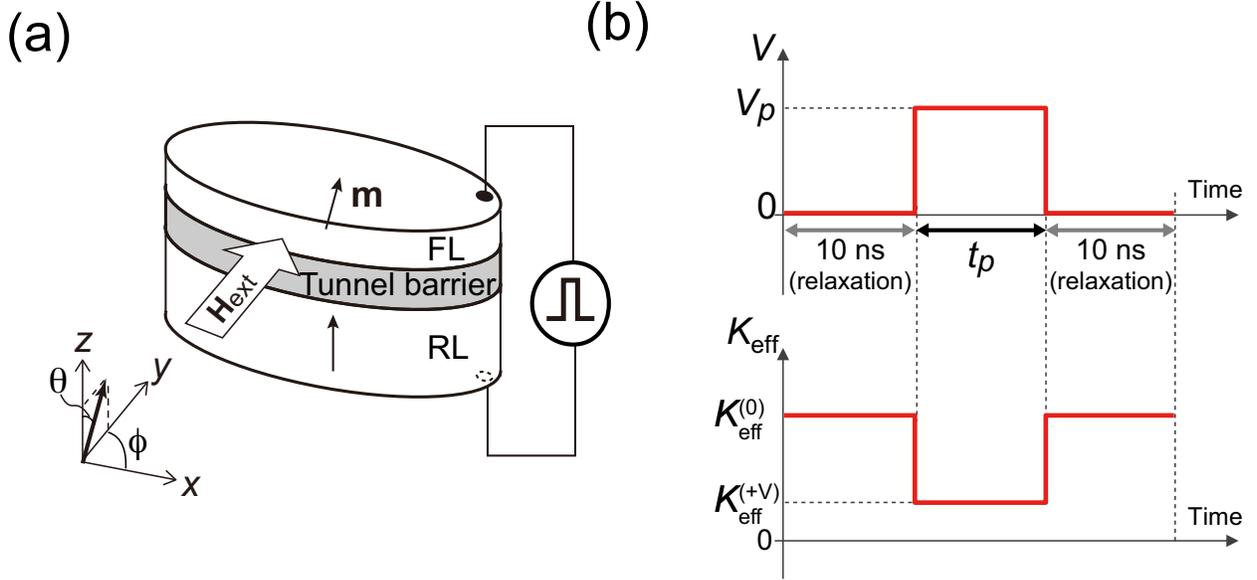}%
  \caption{
  \label{fig:fig1}
  (a) An elliptical-cylinder magnetic tunnel junction (MTJ), with an external magnetic field ($\textbf{H}_{\rm ext}$), 
  and the definitions of the Cartesian coordinates $(x, y, z)$.
  FL and RL denote the free layer and the reference layer, respectively. 
  The $x$ axis is parallel to the major axis of the ellipse, and the external field, $\textbf{H}_{\rm ext}$, is applied in the positive $y$ direction, 
  which is parallel to the minor axis of the ellipse.
  (b) Top: the shape of voltage pulse: the amplitude and duration of the pulse are $V_{p}$ (positive value) and $t_{p}$, respectively.
  Bottom: the corresponding time dependence of the effective anisotropy constant $K_{\rm eff}$: at $V=0$, 
  it takes the value $K_{\rm eff}^{\rm (0)}$. When $V=V_{p}$, $K_{\rm eff}=K_{\rm eff}^{\rm (+V)}$.
  }
\end{figure}

In dynamic precessional switching, the pulse width after half a precession period must be controlled precisely to obtain a low write-error rate (WER). 
For example, to obtain a WER less than $10^{-3}$, which is the highest WER acceptable for AI image recognition \cite{yeoh_jsap_2022}, 
the pulse width must be controlled on a subnanosecond basis \cite{shiota_evaluation_2016, shiota_reduction_2017,yamamoto_improvement_2019}.
From a practical point of view, however, it is difficult to precisely control the pulse width for all memory cells in a highly integrated circuit because of the distribution of the precession period among the memory cells.

We have previously proposed a writing scheme based on heavily damped precession of the magnetization, 
where the WER is less sensitive to the pulse width \cite{matsumoto_voltage-induced_2019, matsumoto_methods_2019, matsumoto_low-power_2020,one_route_2021}. 
Even when the voltage is applied for over half a precession period, the magnetization is kept near the opposite direction of the initial state, 
so the pulse width does not need to be controlled precisely. This prolonged tolerance of the pulse width is caused by the fast energy dissipation through damping torque during the precession \cite{maruyama_large_2009}.
The WER demonstrated in Refs. \cite{matsumoto_voltage-induced_2019, matsumoto_methods_2019} is on the order of $10^{-4}$, which can be used for AI image recognition. However, the WER needs to be improved further to broaden the application areas of the VCMRAM.


In this paper, we theoretically investigate heavily damped precessional switching 
in an elliptical-cylinder voltage-controlled MTJ  
\cite{deng_ultrafast_2017, miriyala_influence_2019}
under an external magnetic field parallel to the minor axis of the ellipse
using the macrospin model.  We derive the conditions of the anisotropy constant during the voltage pulse and the magnitude of the external field to switch the magnetization. 
We also perform numerical simulations and show that the WER for the elliptical cylinder can be several orders of magnitude smaller than that for the circular cylinder  if the external magnetic field is applied parallel to the minor axis of the ellipse. Detailed analyses based on the numerical simulations reveal that the reduction of the WER is due to the demagnetization field narrowing the magnetization distribution perpendicular to the plane direction immediately before the voltage is applied.

The rest of the paper is organized as follows. 
Section II introduces the theoretical model. 
In Section III,
we show that the WER for the elliptical-cylinder MTJ can be several orders of magnitude lower than that of the circular-cylinder MTJ.
Section IV presents the detailed analysis for determining the optimal conditions for the low WER. 
In Sec. V, we investigate the cause of the reduction in the WER.

\section{theoretical model}
The system we consider is schematically shown in Fig. \ref{fig:fig1}(a). The lateral size of the voltage-controlled MTJ is assumed to be so small that the magnetization dynamics can be described by the macrospin model. The direction of magnetization in the FL is represented by the unit vector ${\bf m} =$ $(m_{x}$, $m_{y}$, $m_{z}) = $ $(\sin \theta \cos \phi$, $\sin \theta \sin \phi$, $\cos \theta$), where $\theta$ and $\phi$ are the polar and azimuthal angles, respectively. 
The $x$ axis is parallel to the major axis of the ellipse. The external in-plane (IP) magnetic field (${\bf H}_{\rm ext}$) is applied parallel to the $y$ axis, so the equilibrium azimuthal angle in the absence of the voltage pulse is $\phi^{(0)} > 0$. Hereafter, the superscript ''(0)'' indicates the quantities at zero bias voltage. The magnetization in the reference layer is fixed to align in the positive $z$ direction.

The energy density of the FL is given by \cite{stiles_spin-transfer_2006}
\begin{align}
  \label{eq:E}
  {\cal E} (m_{x}, m_{y}, m_{z})
  =
   &
  \frac{1}{2} \mu_{0} M_{s}^{2}
  ( N_{x} m_{x}^{2} + N_{y} m_{y}^{2} + N_{z} m_{z}^{2} )
  \nonumber \\
   &
  + K_{u} (1-m_{z}^{2})
  - \mu_{0} M_{s} \textbf{m} \cdot \textbf{H}_{\rm ext},
\end{align}
where the demagnetization coefficients, $N_{x}$, $N_{y}$, and $N_{z}$, are assumed to satisfy $N_{z} \gg N_{y} > N_{x}$. $\mu_{0}$ is the vacuum permeability and $M_{s}$ is the saturation magnetization of the FL. The index of the IP shape-anisotropy field is given by $H_{k}^{\rm (IP)}=M_{s}(N_{y} - N_{x})$ \cite{matsumoto_critical_2016}. $K_{u}$ is the uniaxial anisotropy constant. 
The value of $K_{u}$  can be controlled by applying a bias voltage, $V$, through the VCMA effect, as shown in Fig. \ref{fig:fig1}(b). $K_{\rm eff}$ represents the effective anisotropy constant $K_{\rm eff} = K_{u} - (1/2)  \mu_{0} M_{s}^{2} ( N_{z} - N_{x} )$, and $K_{\rm eff}^{\rm (+V)}$ indicates the value of $K_{\rm eff}$ during the voltage pulse.

The magnetization dynamics are simulated using
the following Langevin equation \cite{brown_thermal_1963}:
\begin{equation}
  \label{eq:Langevin}
  (1+ \alpha^{2}) \frac{{\rm d} {\bf m}}{{\rm d}t}
  = -\gamma_{0} {\bf m}\times
  \left\{
  \left(\bf{H}_{\rm eff} + \bf{h}\right)
  +\alpha
  \left[
    \bf{m}\times\left(\bf{H}_{\rm eff} + \bf{h}\right)
    \right]
  \right\},
\end{equation}
where $t$ is time, $\gamma_{0}$ is the gyromagnetic ratio, and $\alpha$ is the
Gilbert-damping constant.
$\bf{h}$ represents the thermal-agitation field satisfying the
following relations:
$\langle h_{\iota}(t)\rangle=0$ and
$
  \langle
  h_{\iota}(t)h_{\kappa}(t')
  \rangle
  = \left[2\alpha k_{\rm B} T / \left( \gamma_{0} \mu_{0} M_{s} \Omega \right)\right]\delta_{\iota\kappa}\delta(t-t')
$, where $\langle \rangle$ represents the statistical mean,  $\iota,\kappa=x,y,z$,
$k_{\rm B}$ is the Boltzmann constant, $T$ is the temperature,
$\Omega$ represents the volume of the FL, and
$\delta_{\iota\kappa}$ is Kronecker's delta.
${\bf H}_{\rm eff}$ is the effective magnetic field, defined as
\begin{align}
  {\bf H}_{\rm eff}= -\frac {1}{\mu_{0} M_{s}} \frac{\partial}{\partial {\bf m}} {\cal E}.
\end{align}

The initial state of the simulation is prepared by relaxing the magnetization from the equilibrium direction on the upper hemisphere ($m_{z} > 0$) at $K_{\rm eff}=K_{\rm eff}^{(0)}$ for 10 ns. Then, the magnetization dynamics are calculated while applying the voltage pulse for a duration of $t_{p}$.
During the pulse, $K_{\rm eff}$ is reduced to $K_{\rm eff}^{\rm (+V)}$ through the VCMA effect  as shown in Fig. \ref{fig:fig1}(b). 
After the pulse, the anisotropy constant rises to the initial value of $K_{\rm eff} = K_{\rm eff}^{(0)}$. 
The success or failure of switching is determined by the sign of $m_{z}$ after 10 ns of relaxation.

\section{RESULTS}

\subsection{Magnetization dynamics in heavily damped precessional switching}
\begin{figure}[t]
  \includegraphics [width=1\columnwidth] {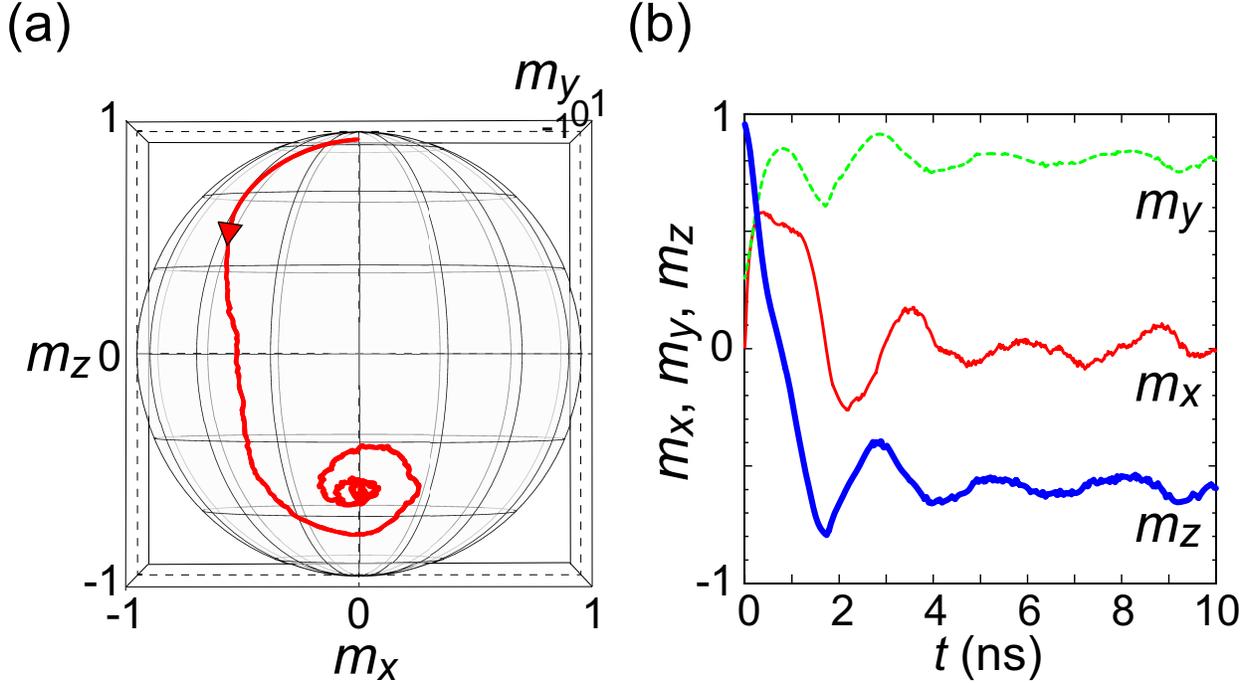}
  \begin{center}
    \caption{
      \label{fig:fig2}
      (a) A typical example of a magnetization trajectory during heavily damped precessional switching
      in the FL of the elliptical-cylinder MTJ at $T= 300$ K.
      (b)
      The temporal evolution of $m_{x}$, $m_{y}$, and $m_{z}$.
    }
  \end{center}
\end{figure}

 Figure \ref{fig:fig2}(a) shows a typical example of a magnetization trajectory during heavily damped precessional switching
in the FL of the elliptical-cylinder MTJ at $T=300$ K. 
We assume that $H_{\rm ext}=400$ Oe,
$M_{s}= 1400$ kA/m,
$\alpha=0.20$, $K_{\rm eff}^{(0)}=70$ kJ/m$^{3}$, and $K_{\rm eff}^{\rm (+V)}=10$ kJ/m$^{3}$. 
The volume of the elliptical FL is assumed to be $\Omega=\pi r_{x}  r_{y} d=123150$ nm$^{3}$, where $r_{x}$  ($r_{y}$) is
half the length of the major (minor) axis of an ellipse
with the aspect ratio  $AR=r_{x}/r_{y}=3$,
and $d=2$ nm is the thickness of the FL.
The demagnetizing constants of the FL are $N_{x}= 0.00535$, $N_{y}=0.02574$, $N_{z}=0.96891$
\cite{beleggia_demagnetization_2005}, which give an IP anisotropy field of
$H_{\rm k}^{\rm (IP)}=359$ Oe.

The magnetization dynamics of heavily damped precessional switching in the elliptical FL are qualitatively the same as those in the circular FL \cite{matsumoto_voltage-induced_2019, matsumoto_methods_2019}. 
Starting from the initial state on the upper hemisphere ($m_{z}>0$), 
the magnetization precesses around the external magnetic field and relaxes toward the equilibrium direction in the lower hemisphere ($m_{z}<0$). 
The temporal evolution of $m_{x}$, $m_{y}$, and $m_{z}$ is shown in Fig. \ref{fig:fig2}(b). It takes less than 2 ns to minimize $m_{z}$. After $m_{z}$ is minimized, 
the magnetization does not return to the upper hemisphere but precesses around the energy minima on the lower hemisphere. 
When the voltage is turned off at any time after 2 ns, the magnetization relaxes toward the equilibrium direction to complete switching.

\subsection{Comparison of WER between circular and elliptical-cylinder MTJs}
\begin{figure}[t]
  \includegraphics [width=0.8\columnwidth] {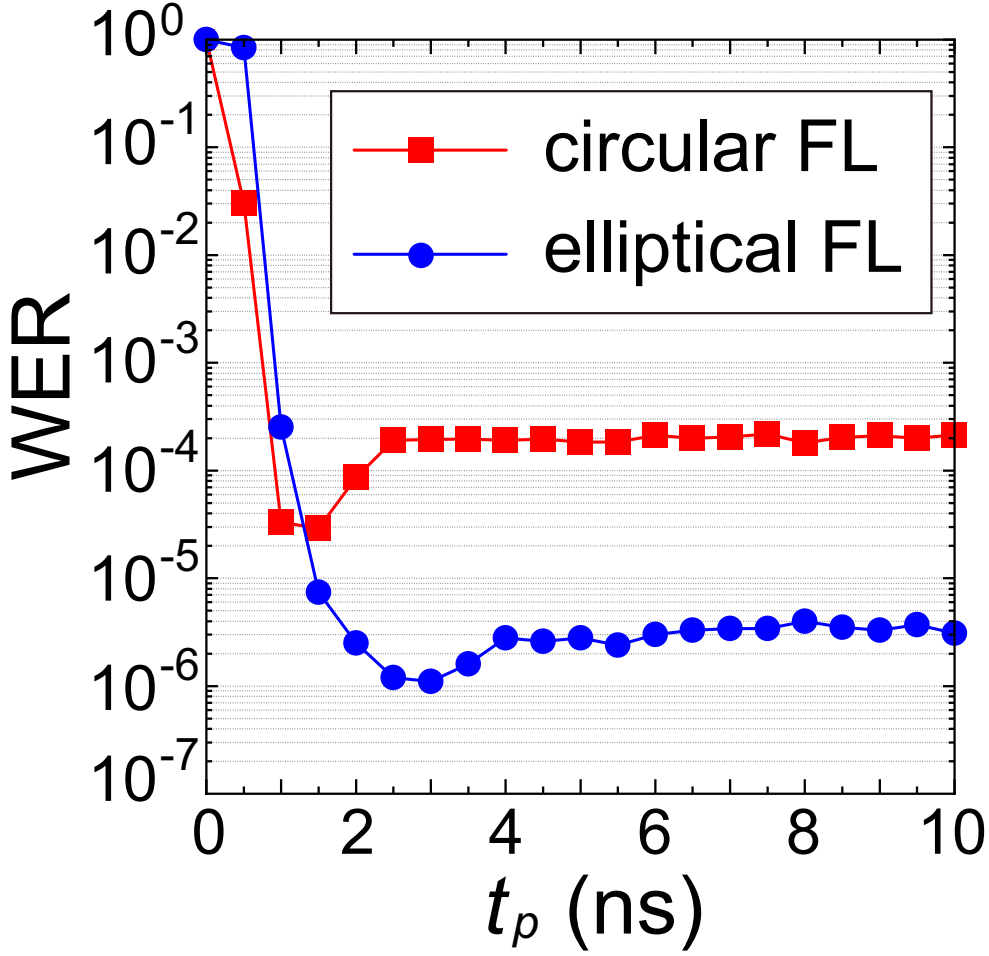}
  \begin{center}
    \caption{
      \label{fig:fig3}
      The $t_{p}$ dependence of the write-error rate (WER). The red squares connected by red lines represent results for the circular FL. 
      The blue circles connected by blue lines represent results for the elliptical FL.
    }
  \end{center}
\end{figure}

The WER of heavily damped precessional switching depends strongly on the shape of the MTJ. 
Figure \ref{fig:fig3} shows the WER of the elliptical-cylinder MTJ (blue) and that of the circular-cylinder MTJ (red) reported in Ref. \cite{matsumoto_voltage-induced_2019}. 
At $t_{p}=10$ ns, the WER of the elliptical-cylinder MTJ is $3.1\times 10^{-6}$ 
which is about 2 orders of magnitude lower than that of the circular-cylinder MTJ ($2.1\times 10^{-4}$).

All of the parameters of the elliptical-cylinder MTJ are the same as those in Fig. \ref{fig:fig2}. 
The volume of the FL of the circular-cylinder MTJ is the same as that of the elliptical-cylinder MTJ, 
i.e., $\Omega=\pi r_{x}  r_{y} d=123150$ nm$^{3}$ with $r_{x}=r_{y}=140$ nm and $d=2$ nm, which is also the same as the FL in Ref. \cite{matsumoto_voltage-induced_2019}. The demagnetization coefficients of the circular FL are $N_{x}=N_{y}=0.01325$, $N_{z}=0.97350$ \cite{beleggia_demagnetization_2005} which yield $H_{\rm k}^{\rm (IP)}=0$ Oe, $\alpha=0.17$, and $K_{\rm eff}^{\rm (+V)}=33$ kJ/m$^{3}$. The other parameters are the same as those of the elliptical FL.

Considering the temperature increase in some computing systems
\cite{chen_advances_2010},
we calculate the WER at $t_{\rm p}=10$ ns and the temperature as 80 $^{\circ}$C ($T= 353$ K).
In the circular-cylinder MTJ,
the WER at $t_{\rm p}=10$ ns 
is $6.1\times10^{-4}$.
In the elliptical-cylinder MTJ,
the WER at $t_{\rm p}=10$ ns 
is $1.9\times10^{-5}$.
In both MTJs,
the WER at $t_{\rm p}=10$ ns increases, but 
the WER is still less than $10^{-3}$ \cite{yeoh_jsap_2022}.

We also conduct simulations adding the pulse-rise time ($t_{\rm r}$)
and the pulse-fall time ($t_{\rm f}$) \cite{yamamoto_write-error_2019}
to the parameters used in Fig. \ref{fig:fig3}.
In the circular-cylinder MTJ,
the WER at $t_{\rm p}=10$ ns 
is insensitive to the introduction of $t_{\rm r}=70$ ps,
but it increases to $1.7\times10^{-2}$ at $t_{\rm r}=200$ ps.
In the elliptical-cylinder MTJ,
the WER at $t_{\rm p}=10$ ns 
is insensitive to the introduction of $t_{\rm r}=40$ ps,
but it increases to $1.3\times10^{-2}$ 
at $t_{\rm r}=200$ ps.
To $t_{\rm r}$, the elliptical-cylinder MTJ is more sensitive 
than the circular-cylinder MTJ.
To $t_{\rm f}$, for both the circular-cylinder MTJ
and the elliptical-cylinder MTJ, 
the WER is insensitive even at $t_{\rm f}=1$ ns.

Note that, in practice, including the external IP magnetic field, 
which is perpendicular to the IP shape-anisotropy field,
may be challenging on a chip.
Competition between the fields can lead to nonuniform
static distribution of the magnetization within the bit.
In addition, the large size assumed in Figs. \ref{fig:fig2} and \ref{fig:fig3} may make the switching nonuniform and the
dynamics might be far from a single-domain precession as considered in
the model. Thus, we conduct micromagnetic simulations, 
and the results are described in Appendix A.
The results support the validity of our analyses.

Even in the case of smaller size, there remain technological challenges.
The elliptical geometry is difficult to scale to small bit
dimensions and increases bit-to-bit variations compared to the
circular shape.

\section{Detailed analyses of magnetization dynamics}
Before investigating the cause of the reduction in the WER, we conduct detailed analyses of the magnetization dynamics in the elliptical FL. 
To save computational time, we analyze the smaller system with $\Omega = S \times  d =15708$ nm$^{3}$. Regardless of $AR$ ranging from 1 to 15, the area and the thickness of the FL are assumed to be $S= 50^{2} \pi$ nm$^{2}$ and $d=2$ nm, respectively. Unless otherwise noted, $M_{s}= 1400$ kA/m , $K_{\rm eff}^{(0)}=200$ kJ/m$^{3}$, $AR=5$, and $(N_{x}$, $N_{y}$, $N_{z})=(0.0075,$  0.0745, $0.9180)$ are assumed.

\subsection{Equilibrium direction of magnetization at $T=$ 0 and $V =$ 0}
\begin{figure}[t]
  \includegraphics [width=1\columnwidth] {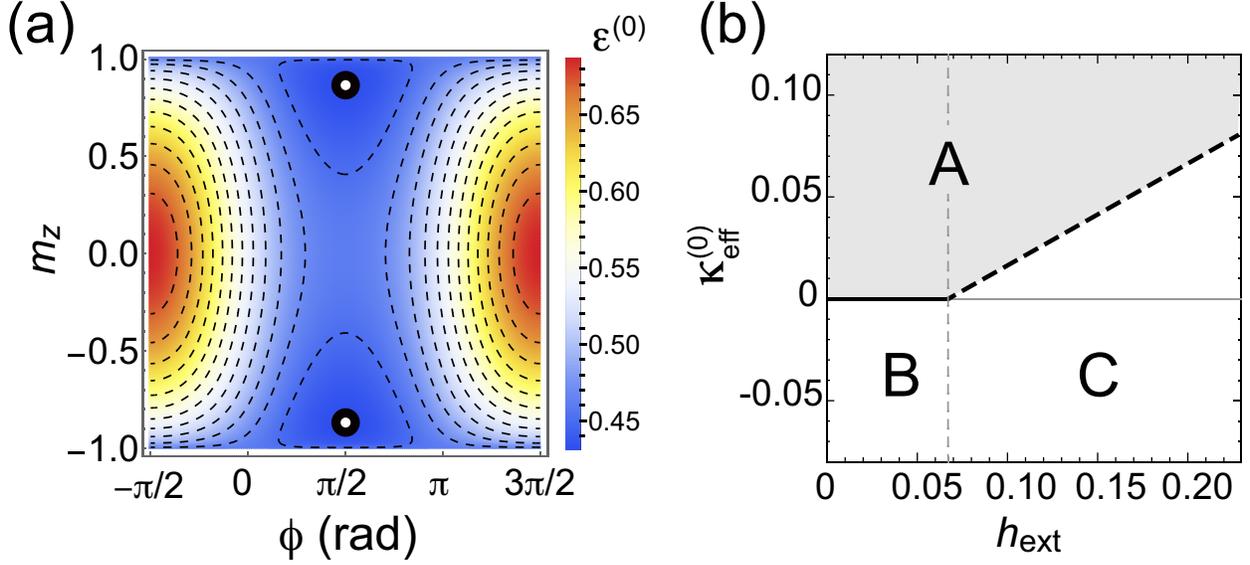}
  \caption{
  \label{fig:fig4}
  (a)
  The energy-density contour plot of
  Eq. (\ref{eq:epsilon}) at 0 V in $\phi - m_{z}$ space. ${\bf m}^{(0)}$ is indicated by open circles.
  (b)
  The classification of $h_{\rm ext}$-$\kappa_{\rm eff}^{(0)}$ space for the calculation of
  ${\bf m}^{(0)}$. In the shaded region {\sf A}, $|m_{z}^{(0)}| > 0$.
  In the white regions, $m_{z}^{(0)} = 0$.
  In region {\sf B}, $0< m_{y}^{(0)} < 1$.
  In region {\sf C}, $m_{y}^{(0)} = 1$, and $m_{x}^{(0)} = 0$.
  In regions {\sf A}, {\sf B}, and {\sf C}, ${\bf m}^{(0)}$ has different analytical
  expressions.
  }
\end{figure}

The equilibrium direction of magnetization at $T=0$ and $V=0$ (${\bf m}^{(0)}$) is obtained by minimizing ${\cal E}$. In this subsection and the next, 
we calculate ${\bf m}^{(0)}$ and analyze the magnetization dynamics using the dimensionless energy density, 
$\varepsilon$, defined as follows \cite{stiles_spin-transfer_2006}:
\begin{align}
  \label{eq:epsilon}
  {\varepsilon} (m_{x}, m_{y}, m_{z})  =
  \frac{1}{2}  ( N_{x} m_{x}^{2} + N_{y} m_{y}^{2} + N_{z} m_{z}^{2} )  \nonumber \\
  + \kappa (1-m_{z}^{2})  - h_{\rm ext} m_{y},
\end{align}
where $\varepsilon={\cal E}/(\mu_{0}M_{s}^{2})$, $\kappa = K_{u}/(\mu_{0}M_{s}^{2})$, and
$h_{\rm ext} = H_{\rm ext} / M_{s}$. 
Without loss of generality, we assume that $h_{\rm ext} > 0$. 
At $V=0$, the dimensionless anisotropy constant is $\kappa=\kappa^{(0)} = K_{u}^{(0)}/(\mu_{0}M_{s}^{2})$. The $\phi$ and $m_{z}$ dependence of ${\varepsilon}^{(0)}$ at $H_{\rm ext}=2000$ Oe is shown in Fig. \ref{fig:fig4}(a), where  ${\bf m}^{(0)} = (m_{x}^{(0)}$, $m_{y}^{(0)}$, $m_{z}^{(0)})=(0$, 0.496, $\pm0.869)$ are indicated by open circles.

To derive the analytical expressions of ${\bf m}^{(0)}$, 
we divide the $h_{\rm ext}$-$\kappa_{\rm eff}^{(0)}$ plane into three regions, {\sf A}, {\sf B}, and {\sf C}, as shown in Fig. \ref{fig:fig4}(b), where $\kappa_{\rm eff} = K_{\rm eff} / (\mu_{0} M_{s}^{2} ) = \kappa - (1/2)   ( N_{z} - N_{x} )$.
In region {\sf A}, indicated by the shaded area, the $z$ component of ${\bf m}^{(0)}$ is nonzero, i.e. $|m_{z}^{(0)}| > 0$. 
In regions {\sf B} and {\sf C}, the magnetization is in the IP direction, i.e. $m_{z} = 0$. 
Therefore, the initial and final state of switching should be in region {\sf A}.

The lower boundary of region {\sf A} is expressed as follows.
For $h_{\rm ext} \leq N_{y} - N_{x}$,
\begin{align}
  \label{eq:LB1A}
  \kappa_{\rm eff} >  0.
\end{align}
For $h_{\rm ext} > N_{y} - N_{x}$,
\begin{align}
  \label{eq:LB2A}
  \kappa_{\rm eff} > \kappa_{\rm eff,c} = \frac{1}{2}(h_{\rm ext} + N_{x} - N_{y} ).
\end{align}
In region {\sf A}, the equilibrium directions of the magnetization are given by
\begin{align}
  \label{eq:mx0A}
  m_{x}^{(0)} & =0,                                                                                                          \\
  \label{eq:my0A}
  m_{y}^{(0)} & = \frac{h_{\rm ext}}{ 2\kappa^{(0)} + N_{y}- N_{z}  } ,                                                            \\
  \label{eq:mz0A}
  m_{z}^{(0)} & = \pm \sqrt{  \frac{ (2\kappa^{(0)} + N_{y} - N_{z} )^{2} - h_{\rm ext}^{2} }{ (2\kappa^{(0)} + N_{y} - N_{z}  )^{2} }}.
\end{align}
By substituting parameters used in Fig. \ref{fig:fig4}(a) into Eqs. (\ref{eq:mx0A}) - (\ref{eq:mz0A}), 
we have ${\bf m}^{(0)} = (m_{x}^{(0)}$, $m_{y}^{(0)}$, $m_{z}^{(0)})=(0$, 0.496, $\pm0.869)$,
which is the same as the result of the numerical calculation.

The boundaries of region {\sf B} are given by
\begin{align}
  \label{eq:BundaryB1}
  h_{\rm ext} \leq N_{y} - N_{x},
\end{align}
and
\begin{align}
  \label{eq:BundaryB2}
  \kappa_{\rm eff} \leq  0.
\end{align}
In region {\sf B}, we have
\begin{align}
  \label{eq:mm0B}
  m_{x}^{(0)} & = \pm \sqrt{1- (h_{\rm ext}/(N_{y} - N_{x}))^{2}}, \\
  m_{y}^{(0)} & =h_{\rm ext}/(N_{y} - N_{x}) ,                     \\
  m_{z}^{(0)} & =0.
\end{align}

The boundaries of region {\sf C} are given by
\begin{align}
  \label{eq:BundaryC1}
  h_{\rm ext} > N_{y} - N_{x},
\end{align}
and
\begin{align}
  \label{eq:BundaryC2}
  \kappa_{\rm eff} \leq \kappa_{\rm eff,c} = \frac{1}{2}(h_{\rm ext} + N_{x} - N_{y} ).
\end{align}
In region {\sf C}, we have
\begin{align}
  \label{eq:mm0C}
  m_{x}^{(0)} & =0, \\
  m_{y}^{(0)} & =1, \\
  m_{z}^{(0)} & =0.
\end{align}

\subsection{Magnetization dynamics at $T=0$}

\begin{figure*}[t]
  \includegraphics [width=1\linewidth] {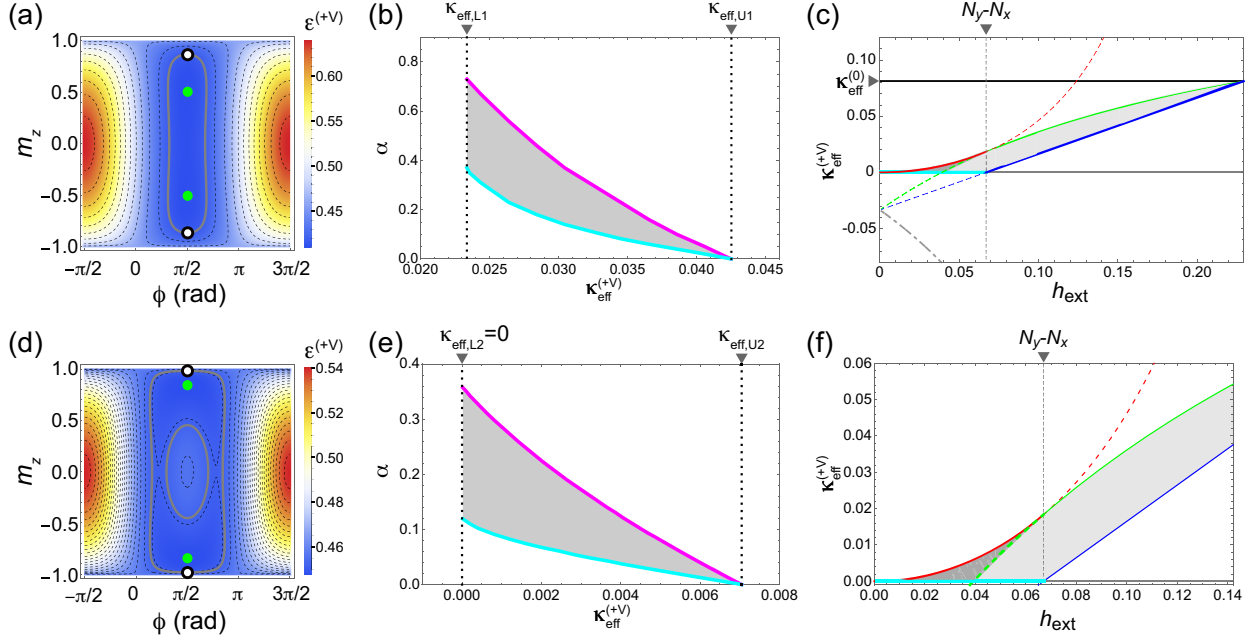}
  \caption{\label{fig:fig5}
  The magnetization dynamics at $T=0$ in the elliptical FL
  with aspect ratio ($AR$) 5 and $K_{\rm eff}^{\rm (0)}=200$ kJ/m$^{3}$ ($\kappa_{\rm eff}^{(0)}=0.0812$).
  (a) The energy-density contour plot of Eq. (\ref{eq:epsilon}) in $\phi - m_{z}$ space 
  during the application of a voltage for $K_{\rm eff}^{(\rm +V)} = 80$ kJ/m$^{3}$ ($\kappa_{\rm eff}^{\rm (+V)}=0.0325$) and $H_{\rm ext}=2000$ Oe ($h_{\rm ext}=0.114 > N_{y} -N_{x}=0.0670$).
  The open circles indicate ${\bf m}^{(0)}$.
  ${\bf m}^{(0)}$ is determined at $\kappa_{\rm eff}^{(0)}$.
  The thick gray dotted curve represents the contour with the same energy
  density as $\varepsilon(\bf{m}^{(0)})$.
  The solid green circles indicate new equilibrium directions (${\bf m}_{\rm eq}^{(\rm +V)}$)
  at $\kappa_{\rm eff}^{(\rm +V)}$.
  (b) The region of heavily damped precessional switching (shaded region) in $\kappa_{\rm eff}^{(\rm +V)} - \alpha$ space for the same parameters as (a).
  (c) The region of heavily damped precessional switching (lighter- and darker-shaded regions) and its boundary in $h_{\rm ext} - \kappa_{\rm eff}^{(\rm +V)}$ space. At given $h_{\rm ext}$, dropping $\kappa_{\rm eff}$ from $\kappa_{\rm eff}^{(0)}$ to $\kappa_{\rm eff}^{\rm (+V)}$ in the shaded regions can induce heavily damped precessional switching at an optimal value of $\alpha$.
  (d)The same plot as in (a) for $K_{\rm eff}^{\rm (+V)}=15$ kJ/m$^{3}$ ($\kappa_{\rm eff}^{\rm (+V)}=0.00609$) and $H_{\rm ext}=750$ Oe ($h_{\rm ext}=0.0426 < N_{y} -N_{x}$).
  (e) The region of heavily damped precessional switching (shaded region) in $\kappa_{\rm eff}^{(\rm +V)} - \alpha$ space for the same parameters as (d).
  (f) An enlarged view of (c).
  }
\end{figure*}

The application of a bias voltage modifies the anisotropy constants from $K_{\rm eff}^{(0)}$ to $K_{\rm eff}^{\rm (+V)}$ and destabilizes the initial state. 
Under the optimal conditions of $K_{\rm eff}^{(0)}$ and $K_{\rm eff}^{\rm (+V)}$, the precessional motion of magnetization around the IP magnetic field is induced \cite{matsumoto_voltage-induced_2018}.
For example, in the case of the elliptical-cylinder MTJ with $AR=5$ and $H_{\rm ext}=2000$ Oe ($h_{\rm ext}=0.114$), a change from $K_{\rm eff}^{(0)}=200$ kJ/m$^{3}$ to $K_{\rm eff}^{\rm (+V)}=80$ kJ/m$^{3}$ (from $\kappa_{\rm eff}^{(0)}=0.0812$ to $\kappa_{\rm eff}^{\rm (+V)}=0.0325$) induces the precession. 
In this case, the contour plot of $\epsilon$ changes from Fig. \ref{fig:fig4}(a) to Fig. \ref{fig:fig5}(a). 
Because the energy contour (gray curve) including ${\bf m}^{(0)}$ (open circles) passes $m_{z}=0$, the magnetization can go down to the lower hemisphere to switch its direction. This condition yields an upper bound of $\kappa_{\rm eff}^{\rm (+V)}$. We label this upper bound as $\kappa_{\rm eff, U1}$, which is indicated by the dotted vertical line in Fig. \ref{fig:fig5}(b).

Note that in the condition of Fig. \ref{fig:fig5}(a), the heavily damped precessional switching is induced at relatively high $\alpha$ ($0.11\leq \alpha \leq 0.30$) while dynamic precessional switching is induced at lower $\alpha$ ($\alpha < 0.11$). 
This is because, as shown in Fig. \ref{fig:fig5}(a), the equilibrium directions of $\bf{m}$ at $\kappa_{\rm eff}^{\rm (+V)}$, ${\bf m}_{\rm eq}^{(\rm +V)}$, indicated by the solid green circles exist on both the upper and the lower hemispheres. 
In such a case, the magnetization can relax to the counterpart ${\bf m}_{\rm eq}^{(\rm +V)}$ after half a precession period even during the application of the bias voltage.

In Fig. \ref{fig:fig5}(b), the values of $(\kappa_{\rm eff}^{\rm (+V)}$ and $\alpha)$ that enable heavily damped precessional switching are indicated by the shaded region. The parameters are $T= 0$, $K_{\rm eff}^{(0)}=200$ kJ/m$^{3}$ ($\kappa_{\rm eff}^{(0)}=0.0812$), $H_{\rm ext}=2000$ Oe ($h_{\rm ext}=0.114$), and $N_{y}-N_{x}=0.0670(< h_{\rm ext})$. 
Similar to the results for the circular MTJ 
reported in Refs. \cite{matsumoto_voltage-induced_2019,matsumoto_methods_2019}, 
the shaded region is triangular. We label its lower bound $\kappa_{\rm eff}^{\rm (+V)}$ as $\kappa_{\rm eff,L1}$. At $\kappa_{\rm eff}^{\rm (+V)} \leq \kappa_{\rm eff,L1}$, heavily damped precessional switching cannot be induced 
because ${\bf m}_{\rm eq}^{(\rm +V)}$ at such $\kappa_{\rm eff}^{\rm (+V)}$ is only located at $m_{z}=0$.

$\kappa_{\rm eff, U1}$ and $\kappa_{\rm eff,L1}$ are analytically calculated in the same way as in Refs. \cite{matsumoto_voltage-induced_2018, matsumoto_voltage-induced_2019} and their $h_{\rm ext}$ dependence is summarized in Figs. \ref{fig:fig5}(c) and (f). Fig. \ref{fig:fig5}(f) is an enlarged view of the low-$h_{\rm ext}$ region in Fig. \ref{fig:fig5}(c). In both the lighter- and darker-shaded regions, heavily damped precessional switching can be induced at appropriate values of $\alpha$.

For $N_{y} - N_{x} <   h_{\rm ext} < 2 \kappa^{\rm (0)} + N_{y} - N_{z} $, the condition on $\kappa_{\rm eff}^{\rm (+V)}$ for the heavily damped precessional switching  is
\begin{align}
  \label{eq:condDampSw1}
  \kappa_{\rm eff,L1}
  <
  \kappa_{\rm eff}^{\rm (+V)}
  <
  \kappa_{\rm eff,U1} .
\end{align}
Here,
\begin{align}
  \label{eq:LB1cond}
  \kappa_{\rm eff,L1} = \frac{1}{2}(h_{\rm ext} + N_{x} - N_{y} ).
\end{align}
This lower bound can be obtained as $\kappa_{\rm eff}$ which yields $m_{z}=0$ in  Eq. (\ref{eq:mz0A}).
$\kappa_{\rm eff,L1}$ is indicated by the solid blue curve in Figs. \ref{fig:fig5}(c) and (f).
This curve is the same as the boundary between regions {\sf A} and {\sf C} in Fig. \ref{fig:fig4}(b).

The upper boundary is
\begin{align}
  \label{eq:myeq1cond}
  \kappa_{\rm eff,U1}
  =
  \frac{ h_{\rm ext} }{ m_{y}^{(0)} + 1}
  - \frac{N_{y} - N_{x} }{2},
\end{align}
where $m_{y}^{(0)}$ is given in Eq. (\ref{eq:my0A}).
$\kappa_{\rm eff,U1}$ is indicated by the solid green curve in Figs. \ref{fig:fig5}(c) and (f).

For $0 <  h_{\rm ext} <N_{y} - N_{x}$,
\begin{align}
  \label{eq:condDampSw2}
  \kappa_{\rm eff,L2}
  <
  \kappa_{\rm eff}^{\rm (+V)}
  <
  \kappa_{\rm eff,U2} .
\end{align}
Here,
\begin{align}
  \label{eq:LB2cond}
  \kappa_{\rm eff,L2} = 0.
\end{align}
This lower bound, $\kappa_{\rm eff,L2}$, is indicated by the solid cyan line in Figs. \ref{fig:fig5}(c) and (f).
This line is the same as the boundary between regions {\sf A} and {\sf B} in Fig. \ref{fig:fig4}(b).
An example of $\kappa_{\rm eff,L2}$ is shown in Fig. \ref{fig:fig5}(e),
where $H_{\rm ext}$ is 750 Oe ($h_{\rm ext}=0.0426 < N_{y}-N_{x}=0.0670$).

The upper bound is
\begin{align}
  \label{eq:discriminant2}
   & \kappa_{\rm eff,U2}
  = -  \frac{1}{2}(N_{z} - N_{x} ) + \nonumber              \\
   & \frac{h_{\rm ext}^2 - 2 h_{\rm ext} m_{y}^{(0)} N_{yx}
  + N_{yx} \left[ N_{z} - N_{x}  - \left( m_{y}^{(0)} \right)^2 (N_{z}-N_{y}) \right]}
  {2 \left[ 1 - \left(m_{y}^{(0)} \right) ^2\right] N_{yx} },
\end{align}
where $N_{yx}=N_{y} - N_{x}$. $\kappa_{\rm eff,U2}$ is indicated by a solid red curve in Figs. \ref{fig:fig5}(c) and (f). An example of $\kappa_{\rm eff,U2}$ is shown in Fig. \ref{fig:fig5}(e).

As seen in Figs. \ref{fig:fig5}(c) and (f), the lower ($\kappa_{\rm eff,L2}$) and upper ($\kappa_{\rm eff, U2}$) bounds of $\kappa_{\rm eff}^{\rm (+V)}$ for the heavily damped precessional switching are different from $\kappa_{\rm eff,L1}$ and $\kappa_{\rm eff, U1}$. 
Note that in the darker-shaded region of Figs. \ref{fig:fig5}(c) and (f), there exist two contours at $\varepsilon = \varepsilon(\bf{m}^{(0)}) $, 
as shown in Fig. \ref{fig:fig5}(d), where $K_{\rm eff}^{\rm (+V)}=15$ kJ/m$^{3}$ ($\kappa_{\rm eff}^{\rm (+V)}=0.00609$), $H_{\rm ext}=750$ Oe ($h_{\rm ext}=0.0426 < N_{y}-N_{x}=0.0670$). 
In Fig. \ref{fig:fig5}(c), the bottom gray dotted-dashed curve shows that  $\kappa_{\rm eff}^{\rm (+V)}$ less than the curve is too low to induce even dynamic switching stably
because the energy contour including ${\bf m}^{(0)}$ does not cross $m_{z}=0$ at such low $\kappa_{\rm eff}^{\rm (+V)}$.

\subsection{Dependence of the WER on $\alpha$ and  $K_{\rm eff}^{\rm (+V)}$}
\begin{figure}[t]
  \includegraphics [width=1\columnwidth] {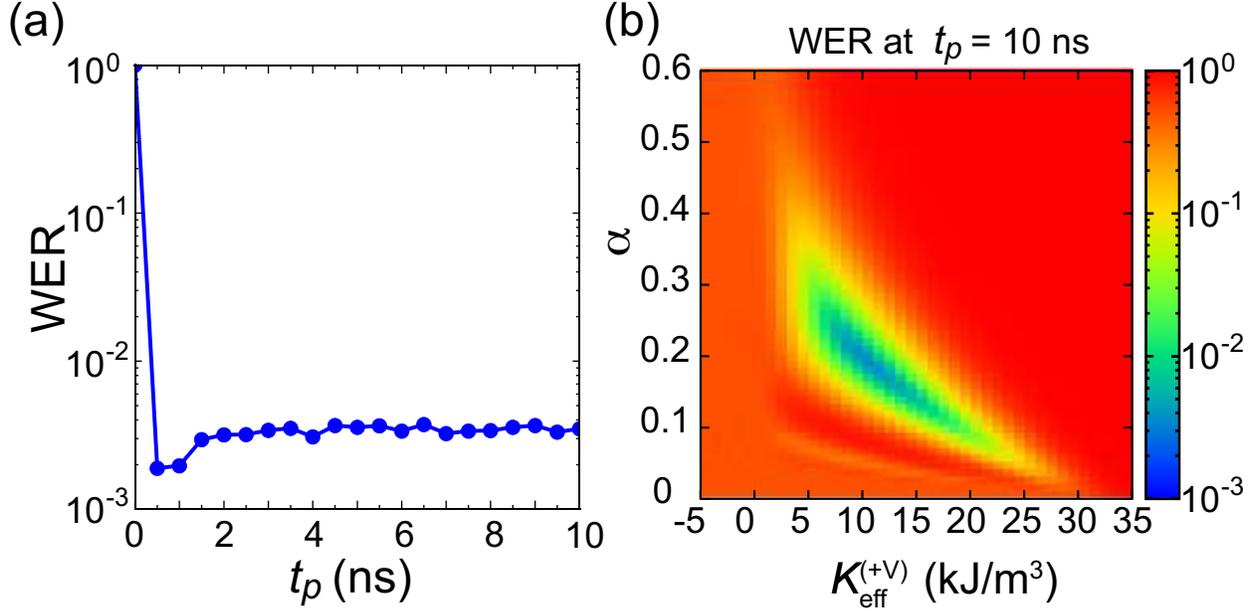}
  \caption{\label{fig:fig6}
    The WER at 300 K and $H_{\rm ext}=1000$ Oe.
    (a) The $t_{p}$ dependence of the WER at
    $K_{\rm eff}^{\rm (+V)}= 10$ kJ/m$^{3}$ and $\alpha=0.19$.
    (b) The $K_{\rm eff}^{\rm (+V)}$ and $\alpha$ dependence of the WER at $t_{p}= 10$ ns.
    The WER is at a minimum value, [WER]$_{\rm min}$, of $3.5 \times 10^{-3}$
    at $K_{\rm eff}^{(\rm +V)}=10$ kJ/m$^{3}$ and $\alpha=0.19$.
  }
\end{figure}

We calculate the WER at $H_{\rm ext}=1000$ Oe ($h_{\rm ext}=0.0568 < N_{y}-N_{x}=0.0670$) and $T=300$ K focusing on the range of $\kappa_{\rm eff}^{\rm (+V)}$ described as Eq. (\ref{eq:condDampSw2}). 
Figure \ref{fig:fig6}(a) shows an example of the $t_{p}$ dependence of the WER calculated in the same way as in Fig. \ref{fig:fig3}. Here, $K_{\rm eff}^{\rm (+V)}= 10$ kJ/m$^{3}$, $\alpha=0.19$, and the other parameters are the same as those in Fig. \ref{fig:fig4}. 
The WER is kept around $3.5 \times 10^{-3}$ for the range of $1.5 \leq t_{p} \leq 10$ ns due to the heavily damped precessional switching.

Figure \ref{fig:fig6}(b) shows the color map of the WER at $t_{p} =10$ ns on the $K_{\rm eff}^{(\rm +V)}$- $\alpha$ plane. 
The WER at $t_{p} =10$ ns is a minimum around the center of the trianglelike region, similarly to Ref. \cite{matsumoto_voltage-induced_2019,matsumoto_methods_2019}. The minimum value of $[{\rm WER}]_{\rm min}=3.5 \times 10^{-3}$ is obtained at $K_{\rm eff}^{(\rm +V)}=10$ kJ/m$^{3}$ and $\alpha=0.19$. 
For example, experimentally, $\alpha$ has been increased by using materials including Pt and Pd
\cite{barman_ultrafast_2007, malinowski_magnetization_2009, mizukami_gilbert_2010, silva_dynamical_2021, bai_data_2012}.

\subsection{Magnetic-field dependence of minimum value of WER}
The minimum value of the WER, [WER]$_{\rm min}$, strongly depends on the magnitude of the external IP magnetic field, $H_{\rm ext}$. 
Because the precession period is inversely proportional to $H_{\rm ext}$, 
the disturbance due to the thermal-agitation field during precession increases as $H_{\rm ext}$ decreases. 
As $H_{\rm ext}$ approaches 0, the WER approaches unity. 
Meanwhile, the energy barrier between the equilibrium directions on the upper and lower hemispheres decreases as $H_{\rm ext}$ increases. 
Above a certain critical value of $H_{\rm ext}$, the WER increases as $H_{\rm ext}$ increases and approaches unity. 
Therefore, there is an optimal value of $H_{\rm ext}$ at which the WER is minimized.

\begin{figure}[t]
  \includegraphics [width=1\columnwidth] {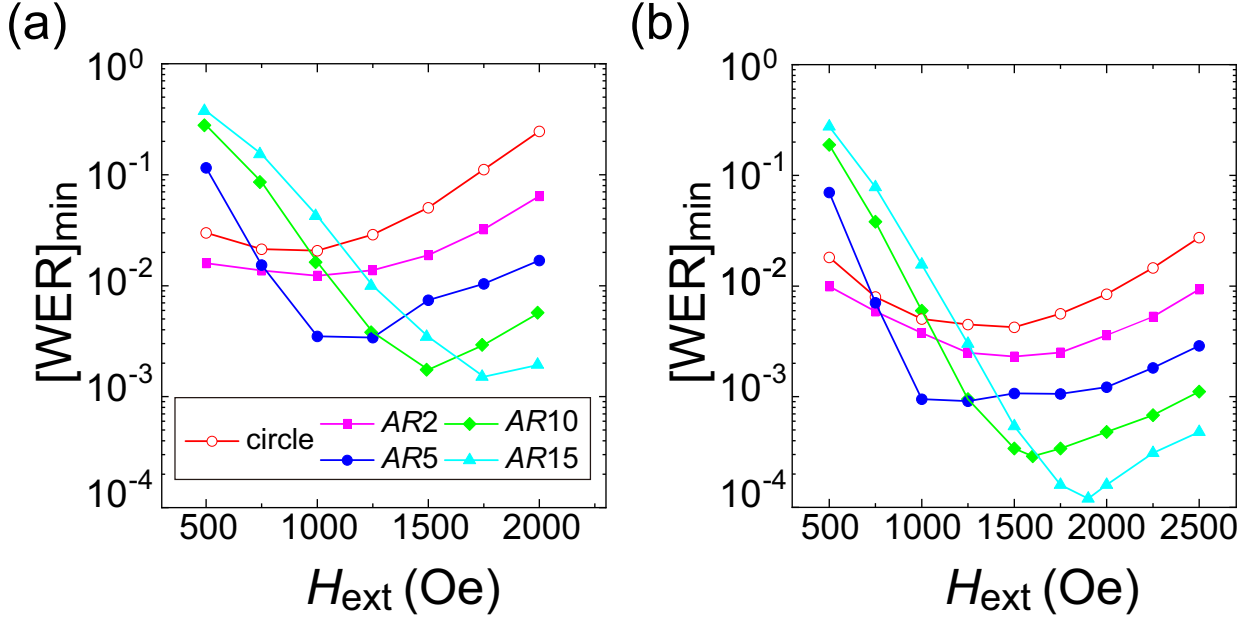}
  \caption{\label{fig:fig7}
    The external in-plane magnetic field dependence of [WER]$_{\rm min}$
    for various $AR$ from 1 (circle) to 15:
    (a) at $K_{\rm eff}^{(0)}=200$ kJ/m$^{3}$ ($0.3 H_{k}^{\rm eff} \approx 900$ Oe);
    (b) at $K_{\rm eff}^{(0)}=300$ kJ/m$^{3}$ ($0.3 H_{k}^{\rm eff} \approx 1300$ Oe).
  }
\end{figure}

To determine the optimal value of $H_{\rm ext}$, we calculate the $H_{\rm ext}$ dependence of $[{\rm WER}]_{\rm min}$ for various values of  $AR$ ranging from 1 (circle) to 15 as shown in Fig. \ref{fig:fig7}(a). 
For the circular MTJ (red open circles), $[{\rm WER}]_{\rm min}$ is minimized around  $H_{\rm ext}=1000$ Oe, where $H_{\rm ext}/H_{k}^{\rm eff}  \approx 0.3$ \cite{matsumoto_methods_2019}.
Here $H_{k}^{\rm eff}=2K_{\rm eff}^{\rm (0)}/\mu_{0} M_{s}$. 
As $AR$ increases, the minimum $[{\rm WER}]_{\rm min}$ decreases and the optimal value of $H_{\rm ext}$ ($H_{\rm ext}^{\rm (opt)}$) increases.

To investigate the effect of the inverse-bias method 
\cite{noguchi_novel_2016,ikeura_reduction_2018,yamamoto_voltage-driven_2020}, 
we conduct similar calculations for a large anisotropy constant $K_{\rm eff}^{(0)}=300$ kJ/m$^{3}$, as shown in Fig. \ref{fig:fig7}(b). 
For each $AR$, the minimum $[{\rm WER}]_{\rm min}$ in Fig. \ref{fig:fig7}(b) is lower than that in Fig. \ref{fig:fig7}(a). 
Also in Fig. \ref{fig:fig7}(b), as $AR$ increases, the minimum $[{\rm WER}]_{\rm min}$ decreases and $H_{\rm ext}^{\rm (opt)}$ increases.

From the results shown in Figs. \ref{fig:fig7}(a) ($0.3 H_{k}^{\rm eff} \approx 900$ Oe) and 7(b) ($0.3 H_{k}^{\rm eff} \approx 1300$ Oe), 
note that $H_{\rm ext}^{\rm (opt)} \approx 0.3 H_{k}^{\rm eff} $ for low $AR (\le 2)$, 
where $M_{s} (N_{y} - N_{x}) \lessapprox 0.3 H_{k}^{\rm eff} $, and $0.3 H_{k}^{\rm eff} < H_{\rm ext}^{\rm (opt)} < M_{s} (N_{y} - N_{x})$ for high $AR (\ge 5)$ 
where $M_{s} (N_{y} - N_{x}) \gtrapprox  0.3 H_{k}^{\rm eff} $. 
The IP demagnetization fields for $AR=2$, 5, 10, and 15 are 
$H_{k}^{\rm (IP)}=516$ Oe, 1178 Oe, 1691 Oe, and 2014 Oe, respectively. These results indicate 
that the increase of the IP demagnetization field causes the reduction in the WER for high $AR$.


The dependence of $[{\rm WER}]_{\rm min}$ on the angles ($\phi_{H}$) 
of $H_{\rm ext}$ is also calculated for $AR=5$, 
$K_{\rm eff}^{(0)}=200$ kJ/m$^{3}$, and $H_{\rm ext}=1000$ Oe.
Here the definition of $\phi_{H}$ is the same as that of 
the $\phi$ illustrated in Fig. \ref{fig:fig1}(a).
In the minor-axis direction, $\phi_{H}=90^{\circ}$, 
$[{\rm WER}]_{\rm min}=3.5 \times 10^{-3}$, as plotted in Fig. \ref{fig:fig7}(a).
At $\phi_{H}=92^{\circ}$, $[{\rm WER}]_{\rm min}$
reaches $[{\rm WER}]_{\rm min}=2.9 \times 10^{-2}$,
which is higher than $[{\rm WER}]_{\rm min}=2.1 \times 10^{-2}$ 
for $AR=1$ (circle), $K_{\rm eff}^{(0)}=200$ kJ/m$^{3}$, 
and $H_{\rm ext}=1000$ Oe.

\section{Effect of IP demagnetization field on the WER }

\begin{figure}[H]
  \includegraphics[width=1\columnwidth]{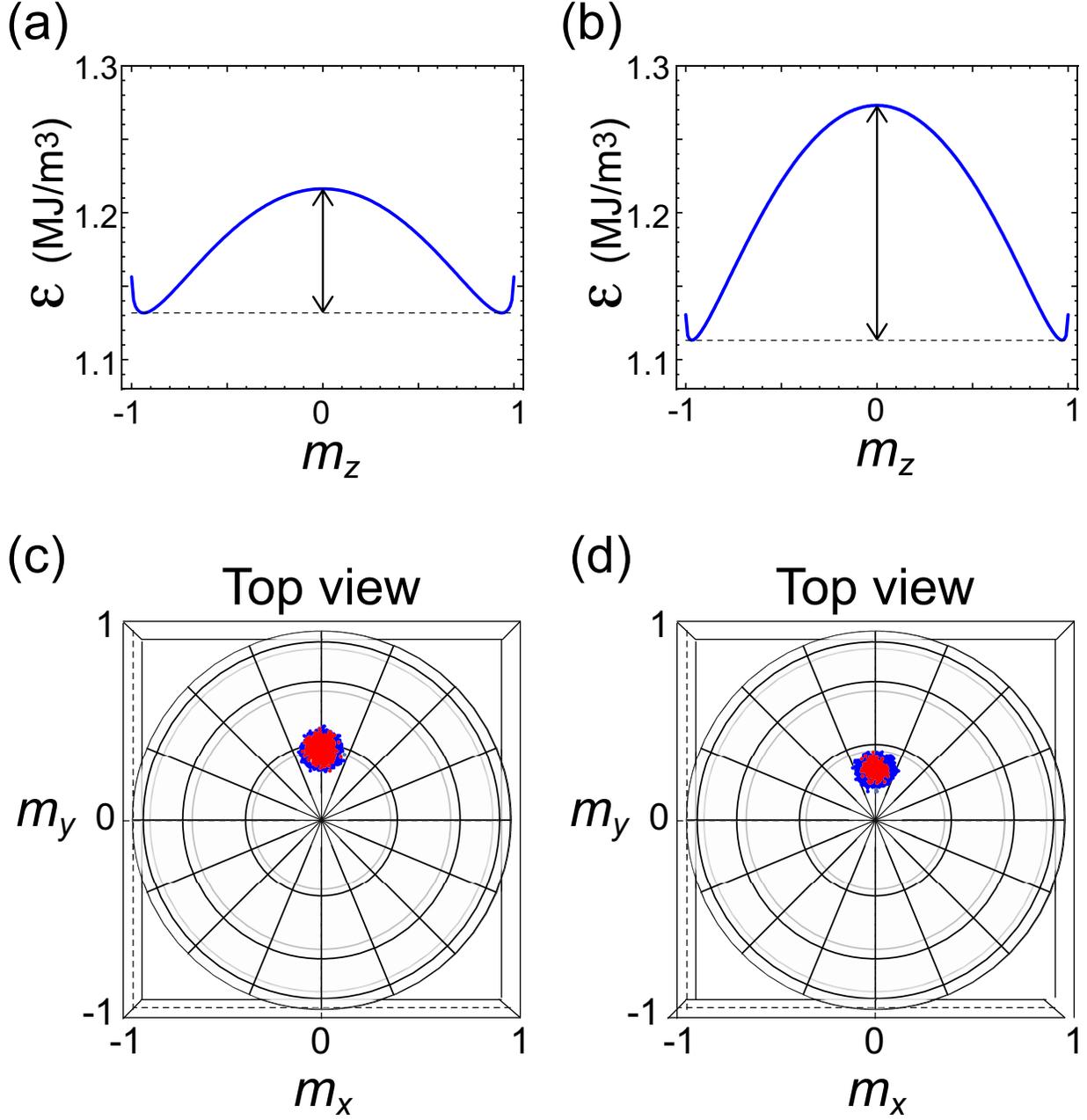}
  \caption{\label{fig:fig8}
    (a) The $m_{z}$ dependence of the energy density (Eq. (\ref{eq:E}))
    at $m_{x}=0$ for the circular FL.
    (b) The same plot for the elliptical FL with $AR=5$.
    (c) The distribution of the magnetization unit vector in the circular FL with $\alpha=0.16$ immediately before application of the voltage pulse, ${\bf m}^{(0) \prime}$, at $T=300$ K and $V=0$.
    $10^{5}$ trials are conducted and
    each blue dot corresponds to ${\bf m}$ after each trial.
    Among the blue dots, the distribution of ${\bf m}^{(0) \prime}$
    that will result in a write error
    after voltage application and subsequent relaxation for 10 ns
    is highlighted as red dots.
    (d) The same plot for the elliptical FL with $AR=5$ and $\alpha=0.19$.
    In all panels, $K_{\rm 1,eff}^{(0)}=200$ kJ/m$^{3}$, and $H_{\rm ext}=1000$ Oe.
    The other parameters are the same as in Fig. \ref{fig:fig6}.
  }
\end{figure}

To analyze the effect of the IP demagnetization field on the WER, 
we compare the $m_{z}$ dependence of the energy density (Eq. (\ref{eq:E})) at $m_{x}=0$, $K_{\rm 1,eff}^{(0)}=200$ kJ/m$^{3}$, and $H_{\rm ext}=1000$ Oe between the circular FL (Fig. \ref{fig:fig8}(a)) and the elliptical FL with $AR=5$ (Fig. \ref{fig:fig8}(b)). 
The other parameters are the same as in Fig. \ref{fig:fig6}. 
The energy-barrier height indicated by the two-headed arrow in Fig. \ref{fig:fig8}(b) is higher than that in Fig. \ref{fig:fig8}(a). 
The value of the energy-barrier height is given in Table \ref{tab:table1}. 
In the elliptical FL, the energy-barrier height is enhanced by the demagnetization energy. 
This enhancement is similar to the enhancement from increasing $K_{\rm 1,eff}^{(0)}$. 
Therefore, the stability of ${\bf m}$ before the application of $V$ is expected to increase 
as $AR$ increases.

\begin{table*}
  \caption{\label{tab:table1}
    A comparison between the circular free layer
    and the elliptical free layer with $AR=5$
    in terms of the energy-barrier height in Figs. \ref{fig:fig8}(a) and (b) and
    the number and distribution of the blue and red dots in Figs. \ref{fig:fig8}(c) and (d).
  }
  \hspace{-3cm}
\scalebox{0.8}{
  \begin{ruledtabular}
    \begin{tabular}{ccccc}
      Geometry of free layer                    & (a) Circle, $\alpha=0.16$, error / all & (b) $AR=5$ ellipse, $\alpha=0.16$,  all & (c) $AR=5$ ellipse, $\alpha=0.19$, error / all \\ \hline
      Barrier height (kJ/m$^{3}$)               & 84.5                                   & 160                                  & 160                                         \\
      Number of dots                            & 2318 (red dots) / $10^{5}$ (blue dots)  & $10^{5}$                             & 350  (red dots)  / $10^{5}$  (blue dots)     \\
      Standard deviation of $m_{z}^{(0)\prime}$ & 0.01173 / 0.00967                      & 0.00559                              & 0.00744 /  0.00559                          \\
      Standard deviation of $\phi^{(0)\prime}$  & 0.0697 / 0.0734                        & 0.1031                               & 0.0987 /  0.1031                            \\
    \end{tabular}
  \end{ruledtabular}
}
\end{table*}

Using the parameters in Figs. \ref{fig:fig8}(a) and (b), the distribution of the initial states (${\bf m}^{(0) \prime}$) at $T=$300 K is compared between the circular FL (Fig. \ref{fig:fig8}(c)) and the elliptical FL with $AR=5$ (Fig. \ref{fig:fig8}(d)). 
The initial states are obtained by relaxing the magnetization from  ${\bf m}^{(0)}$ with $m_{z}^{(0)}>0$ for 10 ns. 
The relaxation is conducted 10$^{5}$ times. ${\bf m}^{(0) \prime}$ after each simulation is plotted by the blue dots in Figs. \ref{fig:fig5}(c) and (d). 
The standard deviation of the distribution in the $m_{z}$ and $\phi$ directions (the standard deviation of $m_{z}^{(0)\prime}$, $\delta_{z}$, and the standard deviation of $\phi^{(0)\prime}$, $\delta_{\phi}$) are also listed in Table \ref{tab:table1}. 
In the elliptical FL, the standard deviation of $m_{z}^{(0)\prime}$ is smaller than that in the circular FL, 
whereas the standard deviation of $\phi^{(0)\prime}$ is not. Note that $\alpha=0.16$ in Fig. \ref{fig:fig8}(c) and $\alpha=0.19$ in Fig. \ref{fig:fig8}(d) 
because each $\alpha$ yields $[{\rm WER}]_{\rm min}$ at $H_{\rm ext}=1000$ Oe in Fig. \ref{fig:fig7}(a).

To clarify the cause of the write error in the heavily damped precessional switching,
we plot ${\bf m}^{(0) \prime}$, which results in the write error as shown by the red dots in Figs. \ref{fig:fig8}(c) and (d). 
The number of red dots and the corresponding standard deviation of $m_{z}^{(0)\prime}$ and  $\phi^{(0)\prime}$ ($\delta_{z}$ and $\delta_{\phi}$) are also listed in Table \ref{tab:table1}. 
In both the circular FL and the elliptical FL, the $\delta_{z}$ of the red dots is larger than that of the blue dots, 
while $\delta_{\phi}$ of the red dots is smaller than that of the blue dots. 
This indicates that the cause of the write error is $\delta_{z}$ rather than $\delta_{\phi}$ in the heavily damped precessional switching.

In the elliptical FL under external in-plane $\textbf{H}_{\rm ext}$, which is parallel to the minor axis of the ellipse, the error in the heavily damped precessional switching is reduced by the suppression of $\delta_{z}$. $\delta_{z}$ is reduced by the energy barrier which is enhanced by the demagnetization energy. 
Note that in the dynamic precessional switching, where the WER is sensitive to $t_{p}$, large $\delta_{\phi}$ also leads to a high WER, 
because large $\delta_{\phi}$ yields the large distribution of optimal $t_{p}$ among all of the trials 
\cite{endo_electric-field_2010, shiota_induction_2012, shiota_pulse_2012,kanai_electric_2012,shiota_evaluation_2016, grezes_ultra-low_2016, shiota_reduction_2017,yamamoto_thermally_2018, yamamoto_improvement_2019,matsumoto_voltage-induced_2018,matsumoto_write_2022}.

It is expected that, in practice, the energy barrier can be enhanced more noticeably 
in smaller FLs. This is because, in larger FLs, the energy barrier is decreased by subvolume activation effects.
Thus, we perform simulations for the smaller FLs with $S= 25^{2} \pi$ nm$^{2}$
and show the results in Appendix B. There, it is confirmed that  
the $H_{\rm ext}$ dependence in the case of $S= 25^{2} \pi$ nm$^{2}$
is qualitatively the same as that in the case of $S= 50^{2} \pi$ nm$^{2}$
shown in Fig. \ref{fig:fig7}.


\section{conclusions}
We theoretically investigate heavily damped precessional switching 
in a perpendicularly magnetized elliptical-cylinder voltage-controlled MTJ. 
We derive analytical expressions of the conditions of the parameters for heavily damped precessional switching. 
The simulations using the Langevin equation show that the WER in the elliptical FL can be several orders of magnitude lower than that in the circular FL. 
From the distribution of the initial magnetization state immediately before a voltage is applied, 
it is revealed that the error in the heavily damped precessional switching is reduced by the suppression of the distribution in the $z$ direction ($\delta_{z}$) 
and $\delta_{z}$ is reduced by the energy barrier, which is enhanced by the demagnetization energy. 
The results provide a guide to designing
high-density VCMRAM for write-error-tolerant applications such as AI image recognition.

\acknowledgements
This work is partly based on results obtained from a project, JPNP16007,
commissioned by the New Energy and Industrial Technology Development Organization (NEDO), Japan.

\begin{appendices}


\section{Micromagnetic simulations}

\label{sec:appendixA}

\begin{figure}[H]
  \includegraphics [width=0.8\columnwidth] {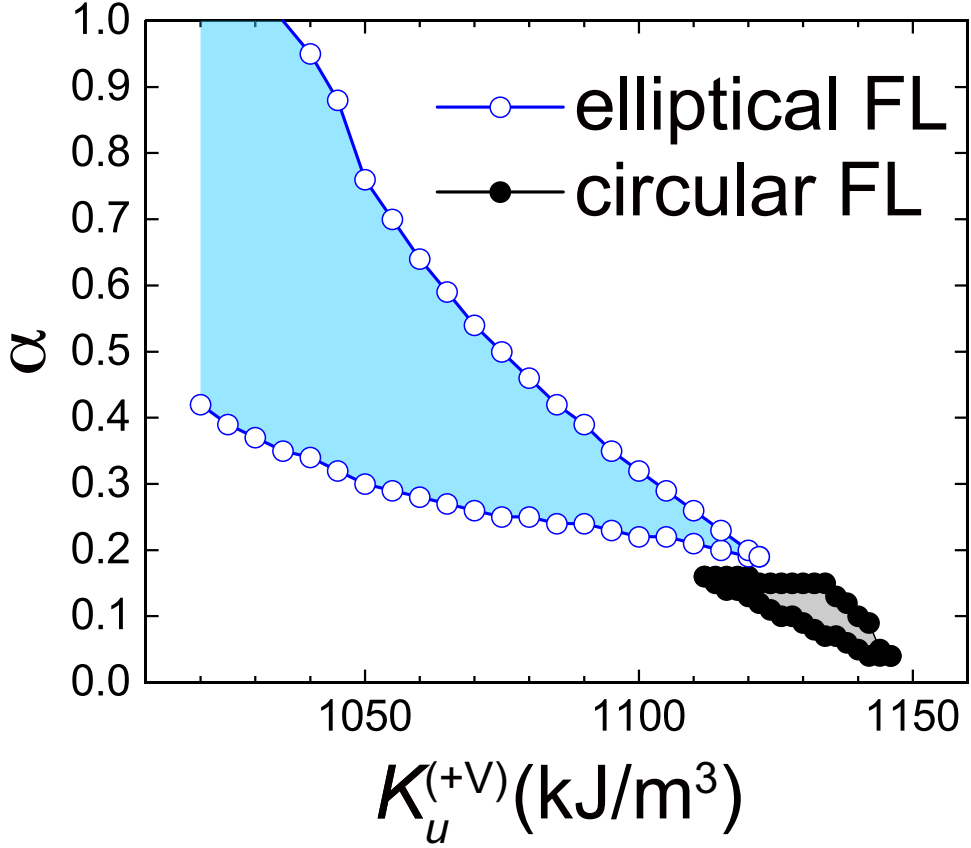}
  \caption{\label{fig:fig9}
 The $K_{\rm eff}^{(\rm +V)} - \alpha$ space diagram 
 obtained by the micromagnetic simulations at 0 K 
 with cell size of 2 nm $\times$ 2 nm and 
 an exchange stiffness constant 
 ($A_{\rm ex}$) of $2 \times 10^{-11}$ J/m.
 $K_{u}= 1252.549$ kJ/m$^{3}$ is assumed and 
 the other parameters are the same as those in Fig. \ref{fig:fig3}.
 The solid black circles on black curves are boundaries 
 for the circular FL (redrawn from Ref. \cite{matsumoto_voltage-induced_2019}).
 The open blue circles on blue curves are boundaries 
 for the elliptical FL.
 The heavily damped precessional switching 
 occurs in the gray and cyan areas.
  }
\end{figure}

We conduct the micromagnetic simulations
by using the MuMax3 software package
\cite{vansteenkiste_design_2014} 
and confirm the $K_{\rm u}^{(\rm +V)} - \alpha$ space diagram 
as shown in Fig. \ref{fig:fig9}.
Here, an exchange stiffness constant 
($A_{\rm ex}$) of $2 \times 10^{-11}$ J/m, 
a cell size of 2 nm $\times$ 2 nm, 
$K_{u}= 1252.549$ kJ/m$^{3}$,
and $T=0$ K are assumed. 
The other parameters are the same as those in Fig. \ref{fig:fig3}.
In the circular FL, the region of 
heavily damped precessional switching (the area in gray)
is quite narrow because of the multimagnetic domains nucleated
during application of the voltage \cite{matsumoto_voltage-induced_2019}.
In the elliptical FL, the region of heavily damped precessional switching (the area in cyan)
is wider than that in the circular FL.
The area in cyan qualitatively agree well with Fig. \ref{fig:fig5}(e)
which is analyzed in the macrospin model.

\section{The WER in smaller junctions}

\label{sec:appendixB}

\begin{figure}[H]
  \includegraphics [width=0.6\columnwidth] {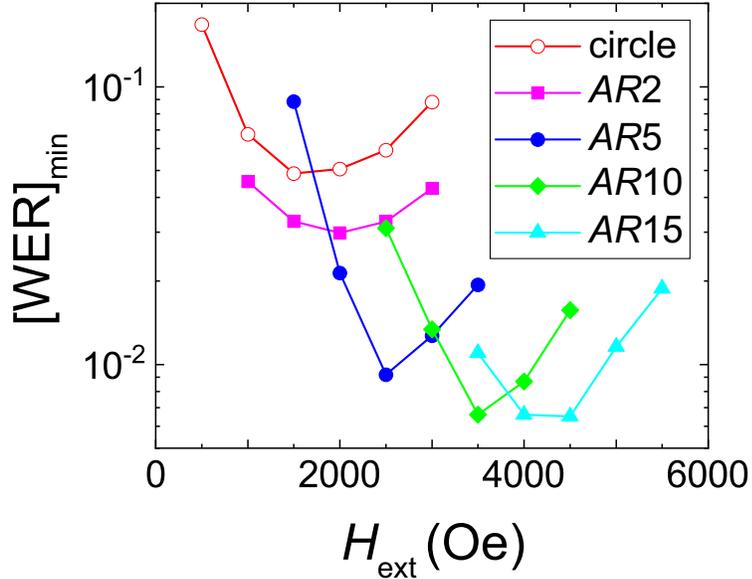}
  \caption{\label{fig:fig10}
    The external in-plane magnetic field ($H_{\rm ext}$) dependence of [WER]$_{\rm min}$
    for various $AR$ from 1 (circle) to 15.
   Here, $K_{\rm eff}^{(0)}=600$ kJ/m$^{3}$ ($0.3 H_{k}^{\rm eff} \approx 1800$ Oe)
   and $M_{s}= 2000$ kA/m.
  }
\end{figure}

Figure \ref{fig:fig10} shows
the dependence of [WER]$_{\rm min}$
on the external in-plane magnetic field ($H_{\rm ext}$) 
in the case of a smaller junction area, $S= 25^{2} \pi$ nm$^{2}$. 
Simulations are conducted in the same way as in Fig. \ref{fig:fig7}. 
Here,
$M_{s}= 2000$ kA/m , $K_{\rm eff}^{(0)}=600$ kJ/m$^{3}$, 
$\Omega = S \times  d =3927$ nm$^{3}$, and $d=2$ nm. 
Regardless of $AR$ ranging from 1 to 15, $S$ and $d$ are the same.

Qualitatively, Fig. \ref{fig:fig10} exhibits the same ($H_{\rm ext}$) dependence as in Fig. \ref{fig:fig7}. 
In the elliptical FLs, [WER]$_{\rm min}$ at optimal $H_{\rm ext}$ ($H_{\rm ext}^{\rm (opt)}$)
is smaller than that in the circular FL.

\end{appendices}


%

\end{document}